\documentstyle[aps,prl,preprint,floats,epsfig]{revtex}

\begin{document}

\preprint{\tighten\vbox{\hbox{\hfil CLNS 97/1497}
                        \hbox{\hfil CLEO 97-16}
}}

\title{\bf  Study of the Decay $\tau^- \to 2\pi^-\pi^+3\pi^0 \nu_\tau$}

\author{CLEO Collaboration}
\date{\today}

\maketitle
\tighten

\begin{abstract}
The decay $\tau^- \to 2\pi^-\pi^+3\pi^0 \nu_\tau$ has been
studied with the CLEO~II detector at the Cornell Electron
Storage Ring (CESR).
The branching fraction is measured to be
$(2.85 \pm 0.56 \pm 0.51)\times 10^{-4}$.
The result is in good agreement with the isospin expectation
but somewhat below the Conserved-Vector-Current (CVC) prediction.
We have searched for resonance substructure in the decay.
Within the statistical precision,
the decay is saturated by the channels
$\tau^- \to \pi^-2\pi^0\omega \nu_\tau$,
$2\pi^-\pi^+\eta \nu_\tau$, and $\pi^-2\pi^0\eta \nu_\tau$.
This is the first observation of this
$\omega$ decay mode and the branching fraction is measured to be
$(1.89^{+0.74}_{-0.67}\pm 0.40) \times 10^{-4}$. 
\end{abstract}
\newpage

{
\renewcommand{\thefootnote}{\fnsymbol{footnote}}
\begin{center}
S.~Anderson,$^{1}$ Y.~Kubota,$^{1}$ S.~J.~Lee,$^{1}$
J.~J.~O'Neill,$^{1}$ S.~Patton,$^{1}$ R.~Poling,$^{1}$
T.~Riehle,$^{1}$ V.~Savinov,$^{1}$ A.~Smith,$^{1}$
M.~S.~Alam,$^{2}$ S.~B.~Athar,$^{2}$
% ...
\end{center}

\small
\begin{center}
$^{1}${University of Minnesota, Minneapolis, Minnesota 55455}\\
$^{2}${State University of New York at Albany, Albany, New York 12222}\\
% ...
\end{center}
\setcounter{footnote}{0}
}
\newpage

The decay of the $\tau$ lepton provides a test of the standard
model prediction of the hadronic weak current.
The decay $\tau^- \to 2\pi^-\pi^+3\pi^0 \nu_\tau$~\cite{conjugate}
is related to $\tau^- \to 3\pi^-2\pi^+\pi^0\nu_\tau$ and
$\tau^- \to \pi^-5\pi^0\nu_\tau$ by isospin symmetry and to the cross
section for $e^+e^- \to 6 \pi$ by the Conserved-Vector-Current (CVC) hypothesis.
The decay can therefore be used to test the isospin and CVC predictions.
However, these predictions are for the $I = 1$ component of the
hadronic weak vector current so any $I = 0$ contribution
to the $e^+e^-$ cross-section or any axial-vector contribution
to the $\tau$ decay through an $\eta$ intermediate state must be
removed before the comparison.
The isospin symmetry also relates the relative branching fractions
among the four possible isospin states~\cite{Pais}, 510 ($4\pi\rho$),
330 ($3\rho$), 411 ($3\pi\omega$), and 321 ($\pi\rho\omega$), which
we denote according to the lowest mass states.
The search for possible $\eta$ and $\omega$ substructure will
therefore be of particular interest.
Any substructure will also be powerful in suppressing hadronic
background in the measurement of the $\nu_\tau$ mass using the
six-pion decay.
Recently, the ALEPH collaboration~\cite{ALEPH} reported substantial
branching fractions for the decays
$\tau^- \to 2\pi^-\pi^+n\pi^0\nu_\tau (n \ge 3)$ and
$2\pi^-\pi^+3\pi^0\nu_\tau$ of
$(1.1\pm 0.4\pm 0.5) \times 10^{-3}$ and
$(2.0\pm 0.6 \pm 0.6) \times 10^{-3}$, respectively.
In this Letter, we present a new result for the branching fraction
of $\tau^- \to 2\pi^-\pi^+3\pi^0 \nu_\tau$ and
the first observation of $\tau^- \to \pi^-2\pi^0\omega \nu_\tau$.
We assume that all three charged particles in the decay are pions.

The data used in this analysis have been collected from $e^+e^-$
collisions at a center-of-mass energy ($\sqrt{s}$) of 10.6~GeV
with the CLEO~II detector at the Cornell Electron Storage Ring (CESR).
The total integrated luminosity of the data sample is $4.68 \rm~fb^{-1}$,
corresponding to the production of $4.27 \times 10^6\ \tau^+\tau^-$ events.   
The CLEO~II detector has been described in detail elsewhere~\cite{Kubota}.

We select events with four charged tracks and zero net charge.
The distance of closest approach of each track to the interaction
point must be within 0.5~cm transverse to the beam and 5~cm along
the beam direction.
The momentum of each track must be greater than $0.02 E_{beam}$
$(E_{beam} = \sqrt{s}/2)$ 
and the polar angle of each track must satisfy $|\cos\theta|<0.90$,
where $\theta$ is the polar angle with respect to the beam.

A photon candidate is defined as a crystal cluster with a minimum
energy of 40~MeV in the barrel region ($|\cos \theta |<0.80$) or
150~MeV in the endcap region ($0.80<|\cos \theta |<0.95$).
The cluster must be isolated by at least 30~cm from the
projection of any charged track on the surface of the calorimeter
unless its energy is greater than 250~MeV.
In addition, the cluster must have a lateral profile of 
energy deposition consistent with that expected for a photon.
A subclass of ``high-quality'' photons is defined to further
discriminate against fake photons; these photons must have a
minimum energy of 150~MeV and pass the isolation requirement
unless their energy exceeds 250~MeV.
All ``high-quality'' photons must be included in the $\pi^0$ reconstruction.
The $\pi^0$ candidates are selected based on a requirement on
$S_{\gamma\gamma} = (m_{\gamma\gamma}-m_{\pi^0})/\sigma_{\gamma\gamma}$,
where $\sigma_{\gamma\gamma}$ is the mass resolution calculated
from the energy and angular resolution of each photon. 
We require all $\pi^0$ candidates to have $-3.5< S_{\gamma\gamma}<2.5$
and be in the barrel; endcap photons are used primarily to
veto background events.

Each event is divided into two hemispheres using the plane
perpendicular to the thrust axis~\cite{thrust}, calculated using both
charged tracks and photons.
There must be one charged track in one hemisphere recoiling
against three charged tracks in the other (1 vs.~3 topology).
In the 1-prong hemisphere, the total invariant mass of charged
tracks and photons must satisfy $M_1 < 1.0 \rm~GeV/c^2$.
We allow up to two high-quality photons in this hemisphere; for the
case of multiple photons, there must be at least one $\pi^0$ candidate. 
For the 3-prong hemisphere, the total invariant mass of charged tracks
and photons must satisfy $M_3 < M_\tau = 1.777 \rm\ GeV/c^2$~\cite{PDG}.  
The magnitude of the total momentum of the particles in the $\tau$ rest
frame, $P^*$, must be less than $0.2 \rm\ GeV/c$.  
In calculating $P^*$, we ignore initial state radiation and 
assume that the $\tau$ direction is the same as the total momentum vector
of the charged tracks and photons in the 3-prong hemisphere. 
This requirement selects events with tau-like kinematics, suppressing
hadronic background and $\tau$ migration background for which the momentum
vector is not a good approximation of the $\tau$ direction. 
The migration background due to photon conversion is further reduced
by a cut on the mass of oppositely charged track pairs ($M_{ee}$),
calculated assuming the electron mass.
In the mass calculation, one of the charged particles must be
identified as an electron, a particle with a shower energy to
momentum ratio in the range, $0.85 < E/P < 1.10$, and, if
available, a measured specific ionization loss that is consistent with
that expected for an electron.
Any event with $M_{ee}<120 \rm \ MeV/c^2$ is rejected.

\begin{figure}[t]
\centering
\centerline{\hbox{\epsfig{figure=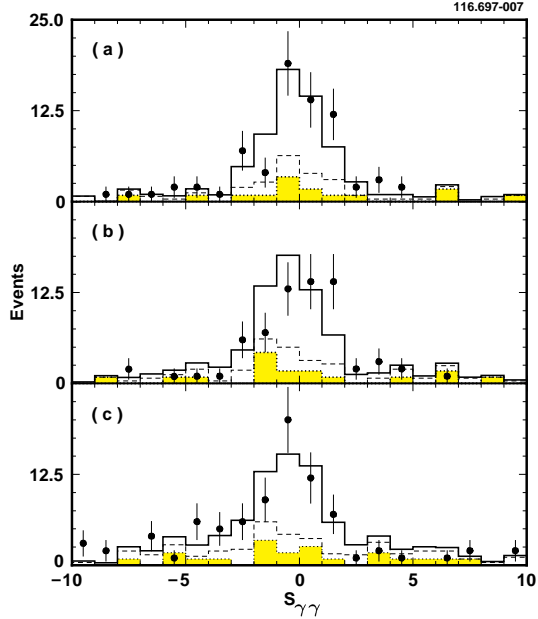,width=2.8in}}}
\vspace{0.10in}
\caption{The $S_{\gamma\gamma}$ distribution for the (a) highest,
(b) intermediate, and (c) lowest energy $\pi^0$ (see text).
The solid histogram is the sum of the signal Monte Carlo and
background (dashed), which includes the $\tau$ migration
and hadronic (shaded) background.} 
\label{figure:S_gg_draft}
\end{figure}

There must be at least six photons in the 3-prong hemisphere,
forming three exclusive $\pi^0$ candidates.
If there is more than one combination that satisfies the requirement,
the one with lowest total
$\chi^2 = \sum^3_{i = 1} [S^2_{\gamma\gamma}]_i$ is selected. 
The $S_{\gamma\gamma}$ distributions of the three $\pi^0$
candidates, classified according to the $\pi^0$ energy, are
shown in Fig.~\ref{figure:S_gg_draft}.
For each distribution, the $S_{\gamma\gamma}$ of the other
two photon pairs must be in their respective signal regions.
An enhancement at zero is evident in all three distributions,
corresponding to the observation of the decay
$\tau^- \to 2\pi^-\pi^+3\pi^0 \nu_\tau$.

\begin{figure}[t]
\centering
\centerline{\hbox{\epsfig{figure=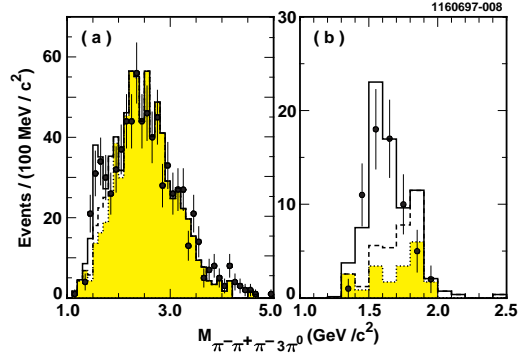,width=2.7in}}}
\vspace{0.1in}
\caption{The invariant mass spectrum of the 3-prong hemisphere
(a) before and (b) after the $P^*$ cut is imposed.
The solid histogram is the sum of the signal Monte Carlo and
background (dashed), which includes the $\tau$ migration
and hadronic (shaded) background.} 
\label{figure:mass_3}
\end{figure}

The detection efficiency and background from $\tau$
migration~\cite{mig} and
hadronic events are calculated with a Monte Carlo (MC) technique.
We use the KORALB/TAUOLA program~\cite{KORALB} for the $\tau$ event
simulation and the Lund program~\cite{Lund} for hadronic events.
The signal decay is modeled using a mixture of
$\tau^- \to 2\pi^-\pi^+\eta\nu_\tau \to 2\pi^-\pi^+3\pi^0\nu_\tau$ (39\%),
$\tau^- \to \pi^-2\pi^0\eta\nu_\tau \to 2\pi^-\pi^+3\pi^0\nu_\tau$ (11\%), and
$\tau^- \to \pi^-2\pi^0\omega\nu_\tau \to 2\pi^-\pi^+3\pi^0\nu_\tau$ (50\%).
The relative mixtures are determined from the measured branching
fractions for the first two decays~\cite{3heta}.
We assume that the $3\pi\eta$ decays proceed through $\pi f_1$ with a spectral function
dominated by the form factor of the $a_1(1260)$ resonance~\cite{Li}.
The $3\pi\omega$ system is modeled using phase space.
The detector response is simulated using the GEANT program~\cite{GEANT}. 
Since the absolute prediction for low-multiplicity hadronic events
may be unreliable, the Lund Monte Carlo is used only to predict the
shape of the 3-prong mass spectrum.
The hadronic background is calculated by normalizing to the number
of data events in the 3-prong hemisphere with $M_3 > 2.0 \rm~GeV$
before the $P^*$ cut is imposed. 
The simulation reproduces the $2\pi^-\pi^+3\pi^0$ invariant mass
spectrum quite well as shown in Fig.~\ref{figure:mass_3}.
An enhancement of events below $M_\tau$ is evident.  
The signal, background, and detection efficiency are summarized
in Table~\ref{table:summary}.
Also listed is the branching fraction extracted after correcting
for the branching fraction of the 1-prong tag of
$(73.0 \pm 0.3)\%$~\cite{PDG}.

\begin{figure}[tb]
\centering
\centerline{\hbox{\epsfig{figure=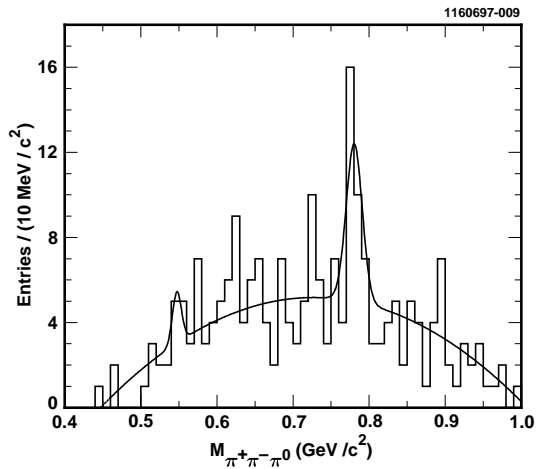,width=2.8in}}}
\vspace{0.1in}
\caption{$\pi^+\pi^-\pi^0$ invariant mass spectrum (6 entries/event).
The curve shows a fit to the data.}
\label{figure:omega}
\end{figure}

The decay $\tau^- \to 2\pi^-\pi^+3\pi^0 \nu_\tau$ can
proceed through different intermediate resonances.
We have recently observed the decays $\tau^- \to
2\pi^-\pi^+\eta\nu_\tau$ via $\eta \to \gamma\gamma$ and
$3\pi^0$ and $\tau^- \to \pi^-2\pi^0\eta\nu_\tau$ via
$\eta \to \gamma\gamma$~\cite{3heta}.
For the latter decay, we also expect an $\eta$ signal in the
$\pi^+\pi^-\pi^0$ mass spectrum.
Figure~\ref{figure:omega} shows the $\pi^+\pi^-\pi^0$ mass
spectrum (6 entries/event), for which the $\pi^0$ candidates
have been kinematically constrained to the nominal $\pi^0$ mass.
To reduce the combinatoric background, we exclude events with a
$3\pi^0$ mass consistent with that of the $\eta$ meson.
There is an enhancement in the $\omega$ mass region corresponding to
the first observation of the decay $\tau^- \to \pi^-2\pi^0\omega\nu_\tau$. 
There is also an indication of a signal for $\eta \to \pi^+\pi^-\pi^0$,
although not statistically significant.
To extract the branching fractions, the spectrum is fitted
using a binned maximum likelihood technique with Gaussians
for the $\omega$ and $\eta$ signals
and a second-order polynomial background.
The $\omega$ and $\eta$ masses and widths have been constrained
to the Monte Carlo expectations. 
The result for $\tau^- \to \pi^-2\pi^0\omega\nu_\tau$
is summarized in Table~\ref{table:summary}. 
If we interpret the small $\eta$ excess as signal, the branching
fraction extracted for $\tau^- \to \pi^-2\pi^0\eta\nu_\tau$
is consistent with the expectation~\cite{3heta}.

\begin{table}[t]
\begin{center}
\caption[]{Summary of signal, background, detection efficiency,
and branching fraction.}
\label{table:summary}
\begin{tabular}{lcc}
Decay Mode          & $2\pi^-\pi^+3\pi^0$ &   $\pi^-2\pi^0\omega$\\ \hline
Data                &       57          & $19.4^{+6.8}_{-6.1}$   \\ 
$\tau$ migration    & $10.2 \pm 2.0$    & $2.3 \pm 0.2$          \\ 
$q\bar q$ background& $8.6  \pm 2.7$    & $0.0^{+1.5}_{-0.0}$    \\
Eff (\%)            & $2.16 \pm 0.07$   & $1.65 \pm 0.09$        \\ \hline
$B(\times 10^{-4})$ & $2.85 \pm 0.56$   & $1.89^{+0.74}_{-0.67}$ \\ 
\end{tabular}
\end{center}
\end{table}

There are several sources of systematic errors as summarized in 
Table~\ref{table:systematics}.
The uncertainty in the photon detection efficiency is
estimated by varying the photon selection criteria.
The systematic error in the decay modeling includes
the uncertainties in the branching fractions of
$\tau^- \to 2\pi^-\pi^+\eta\nu_\tau$ and $\pi^-2\pi^0\eta\nu_\tau$~\cite{3heta}
and the modeling of the $\pi f_1$ and $3\pi\omega$ systems. 
The latter uncertainty is estimated from the difference in the detection efficiency between
the default models and the models based on a $\pi f_1$ phase space and
a $\pi\rho\omega$ resonance with mass of 1600~MeV/c$^2$
and width of 235~MeV/c$^2$.
The uncertainty in the background shape in the $\pi^+\pi^-\pi^0$ mass
spectrum is estimated by using different orders of polynomial.  
As a check of the hadronic background estimate, the branching fraction
for the decay $\tau^- \to 2\pi^-\pi^+3\pi^0\nu_\tau$ has been
measured using a lepton tag and the result is consistent with that for
the non-lepton tag.
Including the systematic errors in quadrature, the final results are:
\begin{eqnarray}
\nonumber
B(\tau^-\to 2\pi^-\pi^+3\pi^0\nu_\tau)&=&(2.85 \pm 0.56 \pm 0.51)\times 
10^{-4}\\
\nonumber
B(\tau^-\to \pi^-2\pi^0\omega\nu_\tau)&=&(1.89^{+0.74}_{-0.67}\pm 0.40)\times 
10^{-4}~,
\end{eqnarray}
where the first error is statistical and the second is systematic.

\begin{table}[bt]
\begin{center}
\caption[]{Summary of systematic errors (\%).}
\label{table:systematics}
\begin{tabular}{lcc}
Mode                 & $2\pi^-\pi^+3\pi^0$ & $\pi^-2\pi^0\omega$ \\ \hline
Luminosity           &         1         &      1            \\ 
Cross-section        &         1         &      1            \\ 
$\tau$ migration     &         5         &      4            \\ 
$q\bar q$ background &         5         &      9            \\ 
Tracking             &         4         &      4            \\ 
Photon eff.          &        15         &     15            \\ 
Decay model          &         4         &      8            \\ 
Fitting              &          -        &      5            \\ 
Eff (MC stat.)       &         3         &      6            \\ \hline
Total                &        18         &     21  
\end{tabular}
\end{center}
\end{table}

\begin{figure}[t]
\centering
\centerline{\hbox{\epsfig{figure=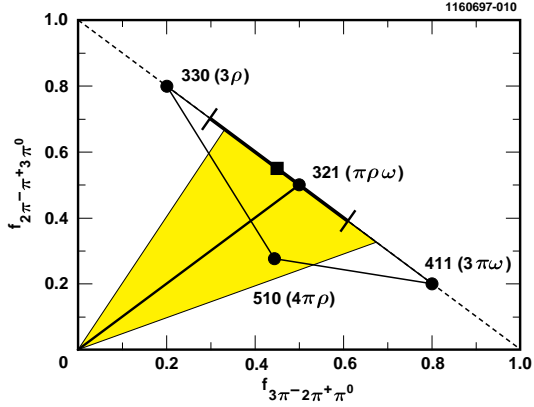,width=2.8in}}}
\vspace{0.1in}
\caption{${\rm f}_{2\pi^-\pi^+3\pi^0}$ vs.~${\rm f}_{3\pi^-2\pi^+\pi^0}$.
The solid line from the origin uses the CLEO~II result on
$B_V(\tau^- \to 2\pi^-\pi^+3\pi^0\nu_\tau)$ and the world
average measurement of
$B_V(\tau^- \to 3\pi^-2\pi^+\pi^0\nu_\tau)$ (see text).
The shaded area indicates the one standard deviation region.
The square with diagonal error bar uses the CLEO~II measurement of
$B(\tau^- \to 2\pi^-\pi^+3\pi^0\nu_\tau)$ from the decay
$\tau^- \to \pi^-2\pi^0\omega\nu_\tau$.
The triangular region indicates the isospin expectation.
Unitarity requires the fractions to be below the dashed line.}
\label{figure:isospin}
\end{figure}

We can test isospin symmetry~\cite{Pais} by comparing the fractions
of the $I = 1$ component of the six-pion decay into
$2\pi^-\pi^+3\pi^0$ and $3\pi^-2\pi^+\pi^0$ states,
\begin{eqnarray}
\nonumber
{\rm f}_{2\pi^-\pi^+3\pi^0} = \frac{B(\tau^-\to 2\pi^-\pi^+3\pi^0\nu_\tau)}
                     {B(\tau^-\to (6\pi)^-\nu_\tau)}
\end{eqnarray}
and similarly for ${\rm f}_{3\pi^-2\pi^+\pi^0}$, where $B(\tau^-\to (6\pi)^-\nu_\tau)$
is the sum of the three branching fractions,
$B(\tau^-\to 2\pi^-\pi^+3\pi^0\nu_\tau)$,
$B(\tau^-\to 3\pi^-2\pi^+\pi^0\nu_\tau)$, and
$B(\tau^-\to \pi^-5\pi^0\nu_\tau)$.
The latter branching fraction has not yet been measured.
However, we expect the branching fraction to be small from CVC using
the measured cross section of $e^+e^- \to 3\pi^+3\pi^-$.
Sobie~\cite{Sobie} and Rouge~\cite{Rouge} have performed
an isospin analysis of these decays and concluded that there
is a discrepancy between the measured decay branching
fractions and the isospin expectation.
However, the authors did not correct for the axial-vector contributions
from the decays $\tau^- \to 2\pi^-\pi^+\eta\nu_\tau$ and
$\pi^-2\pi^0\eta\nu_\tau$.
We correct for these contributions using our recent measurements
of the branching fractions, $(3.5^{+0.7}_{-0.6} \pm 0.7)\times 10^{-4}$
and $(1.4 \pm 0.6 \pm 0.3)\times 10^{-4}$~\cite{3heta}.
Figure~\ref{figure:isospin} shows ${\rm f}_{2\pi^-\pi^+3\pi^0}$
vs.~${\rm f}_{3\pi^-2\pi^+\pi^0}$ with our new measurement of
$B_V(\tau^- \to 2\pi^-\pi^+3\pi^0\nu_\tau)
= (1.41 \pm 0.76)\times 10^{-4}$ and the world average
measurement~\cite{PDG} of
$B_V(\tau^- \to 3\pi^-2\pi^+\pi^0\nu_\tau)
= (1.39 \pm 0.55)\times 10^{-4}$,
where the subscript indicates that this is the vector
component of the branching fraction.
The measurement is in good agreement with the isospin expectation.
Because the $\eta$ and $\omega$ intermediate states saturate
the decay $\tau^- \to 2\pi^-\pi^+3\pi^0\nu_\tau$,
we can also test the isospin prediction using our measurement of
$B(\tau^- \to 2\pi^-\pi^+3\pi^0\nu_\tau)$ from the decay
$\tau^- \to \pi^-2\pi^0\omega\nu_\tau$.
With the assumption that $B(\tau^- \to \pi^-5\pi^0\nu_\tau)$
is negligible, our measurement of
$B_\omega(\tau^- \to 2\pi^-\pi^+3\pi^0\nu_\tau) = (1.68 \pm 0.72)\times 10^{-4}$
is consistent with the isospin expectation
as shown in Fig.~\ref{figure:isospin}.
The result prefers the dominance of the [321] $(\pi\rho\omega)$
isospin state over the [411] $(3\pi\omega)$ state; we are not
able to distinguish the two states from the $\pi\pi^0$ mass spectrum.
Our measurement of $B(\tau^- \to 2\pi^-\pi^+3\pi^0\nu_\tau)$
can also be compared with the CVC predictions by Sobie~\cite{Sobie}
and Eidelman~\cite{Eidelman},
$B(\tau^- \to 2\pi^-\pi^+3\pi^0\nu_\tau) = (1.9 - 7.3) \times 10^{-4}$
and $(2.5 \pm 0.4) \times 10^{-4}$, respectively.
Our result is somewhat below the predictions,
indicating that there could be a substantial $I = 0$ contribution
from $e^+e^- \to \eta\pi^+\pi^-\pi^0$ to
$e^+e^- \to 2\pi^+2\pi^-2\pi^0$ that must be removed before calculating
the CVC predictions.

In conclusion, the decay $\tau^- \to 2\pi^-\pi^+3\pi^0 \nu_\tau$
has been observed.
The branching fraction is in a good agreement with the isospin
expectation~\cite{Sobie,Rouge}, but smaller than the
ALEPH measurement~\cite{ALEPH}.
We have also observed an $\omega$ signal in the six-pion system,
corresponding to the decay $\tau^- \to \pi^-2\pi^0\omega\nu_\tau$.
Within the statistical precision,
the contribution from this decay together with those~\cite{3heta} from
$\tau^- \to 2\pi^-\pi^+\eta\nu_\tau$ and $\pi^-2\pi^0\eta\nu_\tau$
saturate the decay $\tau^- \to 2\pi^-\pi^+3\pi^0 \nu_\tau$.

We gratefully acknowledge the effort of the CESR staff in providing us with
excellent luminosity and running conditions.
This work was supported by 
the National Science Foundation,
the U.S. Department of Energy,
the Heisenberg Foundation,  
the Alexander von Humboldt Stiftung,
Research Corporation,
the Natural Sciences and Engineering Research Council of Canada,
and the A.P. Sloan Foundation.



\begin{thebibliography}{99}

\bibitem{conjugate} In this paper charge conjugate states are implied.

\bibitem{Pais}   A.~Pais, Ann. Phys. {\bf 9}, 548 (1960).
                 For an isospin state $[n_1, n_2, n_3]$, $n_3$ is the number
                 of subsystems of three pions with $I = 0$, $n_2 - n_3$ is
                 the number of subsystems of two pions with $I = 1$, and
                 $n_1 - n_2$ is the number of remaining single pions.

\bibitem{ALEPH}  D.~Buskulic  {\it et al.}, Z. Phys. C {\bf 70}, 579 (1996).

\bibitem{Kubota} Y.~Kubota {\em et~al.}, Nucl. Instrum. Methods A
                 {\bf 320}, 66 (1992).

\bibitem{thrust} E.~Farhi, Phys. Rev. Lett.~{\bf 39}, 1587 (1977).

\bibitem{PDG}    R.~Barnett {\em et~al.}, 
                 Review of Particle Properties, Phys. Rev. D~{\bf 54}, 1 (1996).

\bibitem{mig}    The dominant migration background in the decay
                 $\tau^- \to 2\pi^-\pi^+3\pi^0\nu_{\tau}$
                 ($\pi^-2\pi^0\omega \nu_\tau$) is from
                 $\tau^- \to 2\pi^-\pi^+2\pi^0\nu_{\tau}$
                 ($\pi^-\pi^0\omega \nu_\tau$).

\bibitem{KORALB} S.~Jadach and Z.~Was, Comp.~Phys.~Comm.~{\ \bf 36}, 191 (1985);
                 {\ \it ibid.}~{\ \bf 64}, 267 (1991); S.~Jadach, J.~H.~Kuhn,
                 and Z.~Was, {\ \it ibid.}~{\ \bf 64}, 275 (1991).

\bibitem{Lund}   T.~Sj\"{o}strand and M.~Bengtsson, Comp.~Phys. Comm.~{\ \bf 43},
                 367 (1987).

\bibitem{3heta}  T.~Bergfeld {\it et al.}, CLNS 97/1489 (submitted to Phys. Rev. Lett.).

\bibitem{Li}     B.A.~Li, Phys. Rev. D~{\bf 55}, 1436 (1997).

\bibitem{GEANT}  R.~Brun {\it et al.}, CERN Report No.~CERN-DD/EE/84-1, 1987 
                 (unpublished).

\bibitem{Sobie}  R.J.~Sobie, Z. Phys. C {\bf 69}, 99 (1995).

\bibitem{Rouge}  A.~Rouge, Z. Phys. C {\bf 70}, 65 (1996).

\bibitem{Eidelman} S.I. Eidelman and V.N. Ivanchenko, Nucl. Phys. B (Proc. Suppl.)~{\bf 55C},
                   181 (1997). (Proceedings of the Fourth International
                   Workshop on Tau Lepton Physics, Estes Park, Colorado,
                   1996, edited by J.G. Smith and W. Toki.)

\end{thebibliography}
\end{document}